\documentclass[12pt,a4paper]{JHEP3}
\def\pa{\partial}

\newcommand{\initiate}{\setcounter{equation}{0}}

\newcommand{\beq}{\begin{equation}}
\newcommand{\eeq}{\end{equation}}
\newcommand\be{\begin{equation} }
\newcommand\bea{\begin{eqnarray}}
\newcommand\ee{\end{equation}}
\newcommand\eea{\end{eqnarray}}
\def\ncr{\nonumber\\ }

\def\Tr{{\rm Tr}\,}

\usepackage{epsfig}

\newlength{\myVSpace}% the height of the box
\newcommand\xstrut{\raisebox{-.5\myVSpace}% symmetric behaviour,
  {\rule{0pt}{\myVSpace}}%
}

\def\endtitle{\par\end{quotation}\vskip3.5in minus2.3in\newpage}

\def\a{\alpha}       \def\b{\beta}
         \def\d{\delta}
\def\e{\epsilon}

\def\m{\mu}          \def\n{\nu}
       
       \def\r{\rho}
\def\s{\sigma}

         \def\G{\Gamma}

\def\ca{{\cal A}}

\def\cg{{\cal G}}      
\def\ci{{\cal I}}      
      \def\cl{{\cal L}}

      \def\car{{\cal R}}

\title{The one-loop renormalization of the gauge sector in
  the noncommutative standard model}

\author{
M. Buri\' c , V. Radovanovi\' c
\\Faculty of Physics, University of Belgrade, \\
P. O. Box 368, 11001 Belgrade, Serbia
\\E-mail:
\email{majab@phy.bg.ac.yu}
 \\E-mail:
\email{rvoja@phy.bg.ac.yu}}

\author{
J. Trampeti\'{c} \\Rudjer Bo\v{s}kovi\'{c} Institute, Theoretical Physics 
Division,   \\     P.O.Box 180, 10002 Zagreb, Croatia and \\
 Max Planck Institut f\"{u}r Physik,
F\"{o}hringer Ring 6,
D-80805 M\"{u}nchen, Germany
\\E-mail:
\email{josipt@rex.irb.hr} }

\abstract{
In this paper we construct a version of
the standard model gauge sector on noncommutative space-time
which is one-loop renormalizable
to first order  in the expansion in the noncommutativity parameter $\theta$.
The one-loop renormalizability is obtained
by the Seiberg-Witten redefinition of the noncommutative gauge potential
for the model containing the usual six representations
of  matter fields of the first generation.
}

\begin{document}

\initiate \section{Introduction}

The interest to formulate a consistent quantum field theory
on noncommutative space, besides
from string theory, comes from mathematics  \cite{Kontsevich:1997vb}
and also from phenomenology.
The standard model of elementatry particles (SM)
has been generalized to a noncommutative setting in many
different ways in the literature: the models thus obtained differ
in their physical
properties such as particle content, additional symmetries,
grand unification scheme, etc.
There are two major approaches  to define noncommutative gauge theories.
We use the so-called $\theta$-expanded approach, in which one utilizes the
Seiberg-Witten (SW) map to express noncommutative fields in terms
of physical (commutative) fields \cite{Seiberg:1999vs}.
In this approach, noncommutativity is
treated strictly perturbatively as an expansion in the noncommutativity 
parameter$\theta$. The major advantage is that models
with any gauge group and
any particle content can be constructed
\cite{Wess,Calmet:2001na,Aschieri:2002mc,Goran,Blazenka}.
The action is manifestly gauge invariant; furthermore, it
has been proved that the action is anomaly free
whenever its commutative counterpart is also
anomaly free \cite{Brandt:2003fx}.

There is a number of versions of the noncommutative standard model (NCSM)
in the $\theta$-expanded approach, \cite{Calmet:2001na,Aschieri:2002mc,Goran}.
The argument of renormalizability was previousely not
included in the construction
because it was believed that
field theories on noncommutative Minkowski space
are not renormalizable in general \cite{Wulkenhaar:2001sq, Maja}.
However, a recent positive result on the one-loop renormalizability of
the $\theta$-expanded
noncommutative $\rm SU(N)$ gauge theory opens
different perspectives \cite{Bichl:2001cq,Buric:2005xe}.
The result \cite{Buric:2005xe} is our initial motivation to reexamine
the noncommutative standard model, in particular its gauge sector.
In this paper we show  that it is possible to
construct a version of the NCSM gauge sector
which is one-loop renormalizable up to first order in $\theta$.
To prove renormalizability, we use the freedom in the Seiberg-Witten map.

Another reason to focus on the gauge sector of the NCSM is
the possibility to detect,
in the forthcoming experiments at LHC,  decays which
are forbidden in the SM \cite{Goran,Josip},
like $ Z \to \gamma\gamma$, and/or to find deviations with respect
to the SM-predicted
angular distributions of the differential cross section in
${\bar f} f \to \gamma\gamma$, etc. scatterings
\cite{Hewett:2000zp,Ohl:2004tn}.
In all of these transitions the so-called
triple gauge boson (TGB) couplings contribute. Clearly,
from the perspective of
the safe usage of noncommutativity-induced corrections to the TGB couplings
in further phenomenological analysis of the above processes, it is
important to prove the regular behaviour of these  interactions
with respect to the one-loop renormalizability.
Signatures of noncommutativity in
experimental particle physics were discussed in the literature
from the point of view of collider physics \cite{Abbiendi:2003wv}.
Decays which are strictly forbidden in
the SM by angular momentum conservation and Bose statistics,
known as the Landau-Pomeranchuk-Yang  theorem, as well as
noncommutativity from neutrino astrophysics and neutrino physics
were discussed in  \cite{Goran,Josip} and \cite{Peter}, respectively.

The plan of the paper is the following.
In Section 2 we briefly review the ingredients of the NCSM
relevant to this work.
In Section 3 the renormalizability of the NCSM gauge sector is worked out;
 in Subection
3.3 the counterterms and the final Lagrangian are explicitly given.
Section 4 is devoted to the discussion of the results
and to the concluding remarks.

\initiate \section{ Noncommutative standard model}

\subsection{General considerations}

The noncommutative space which we consider is the flat Minkowski space,
generated by four hermitean coordinates $\widehat x^\mu$
which satisfy the commutation rule
\begin{equation}
[\widehat x^\mu,\widehat x^\nu ] =i\theta^{\mu\nu}={\rm const}.
\label{Mink}
\end{equation}
The algebra of the functions $\widehat\phi (\widehat x)$,
$\widehat \chi (\widehat x)$
on this space can be represented by the algebra of the functions
$\widehat \phi (x)$, $\widehat \chi (x)$ on the commutative
$ {\bf R}^4$ with
the Moyal-Weyl multiplication:
\begin{equation}
\label{moyal} \widehat\phi (x)\star \widehat\chi (x) =
      e^{\frac{i}{2}\,\theta^{\m\n}\frac{\pa}{\pa x^\m}\frac{\pa}{ \pa
      y^\n}}\widehat \phi (x)\widehat \chi (y)|_{y\to x}\ .
\end{equation}
It is possible to represent the action of an arbitrary Lie group $G$
(with the generators denoted by $T^a$) on noncommutative space.
In analogy to the ordinary  case,
one introduces the gauge parameter $\widehat \Lambda ( x)$
 and the vector potential $\widehat V_\m( x)$.  The main difference
 is that the noncommutative
 $\widehat \Lambda$ and $\widehat V_\m $  cannot take values
 in the Lie algebra $\cg$ of the group $G$:  they are
 enveloping algebra-valued.
 The gauge field strength  $ \widehat F_{\mu\nu}$
 is defined in the usual way
\begin{equation}
\widehat F_{\mu\nu} = \pa_\mu\widehat V_\nu -
\pa_\nu \widehat V_\mu - i(\widehat V_\mu\star\widehat V_\nu -
\widehat V_\nu\star\widehat V_\mu ).                         \label{f}
\end{equation}
There is, however, a relation between the noncommutative
gauge symmetry and the commutative one:
it is given by the Seiberg-Witten (SW) mapping \cite{Seiberg:1999vs}.
Namely, the matter fields
 $\widehat \phi$, the gauge fields  $\widehat V_\mu$, $\widehat F_{\mu\nu}$
 and the gauge parameter
$\widehat\Lambda$ can be expanded in the noncommutative  $\theta^{\m\n}$
and in the commutative  $V_{\mu}$ and $F_{\mu\nu}$.
 This expansion  coincides with the expansion in
the generators of the enveloping algebra of $\cg$,
$\{ T^a$, $:T^aT^b:$, $:T^aT^bT^c: \}$;
 here \ $:\ :$ \ denotes the symmetrized product.
The SW map is obtained as a solution to the gauge-closing condition
of   infinitesimal (noncommutative)  transformations.
The  expansions of the NC vector potential and of the field strength,
up to  first order in $\theta$, read
\bea
&&
\widehat V_\r(x) =V_\r(x) -\frac 14 \,\theta ^{\m\n}\left\{ V_\m(x),
\pa _\n V_\r(x) +F_{\n \r}(x)\right\}
+\dots  .
\label{expansion}\\
&&
\widehat F_{\r\s}(x) =   F_{\r\s}(x) +\frac{1}2{}\theta^{\m\n} \{
F_{\m\r}(x),F_{\n\s}(x)\} -\frac{1}{4}\theta^{\m\n}\{ V _\m (x),
(\pa_\n +D_\n )F_{\r\s} (x)\}+\dots
\nonumber\\
\label{FF}
\eea
$D_{\mu}$ is the commutative covariant derivative.

The solution for the SW map given above is  not unique.
As it was shown in
  \cite{Asakawa:1999cu,Wulkenhaar:2001sq},
 along with (\ref{FF}) all expressions  ${\widehat V}^\prime_{\mu}$,
${\widehat F}^\prime_{\mu\nu}$    of the form
\begin{equation}
{\widehat V}^\prime_{\mu} = {\widehat V}_{\mu} +  X_\mu, \quad
{\widehat F}^\prime_{\mu\nu} = {\widehat F}_{\mu\nu} + D_\mu X_\nu - D_\nu X_\mu 
,\label{nonuniq}
\end{equation}
are solutions to the closing condition to linear order, if
 $X_\mu $ is a gauge covariant expression linear in $\theta$, otherwise 
arbitrary. One can think of this transformation as of a redefinition of
 the fields $V_\m$ and $F_{\m\n}$.

Taking the action of the  noncommutative gauge theory
\begin{equation}
S=-\frac{1}{2}\Tr \int d^4x\,\widehat F_{\m\n}\star\widehat F^{\m\n} ,
\label{action}
\end{equation}
and expanding the fields as in (\ref{expansion}-\ref{FF})
and the $\star$-product in $\theta$, we obtain the expression
\begin{equation}
S =-\frac{1}{2}\Tr\int d^4x\,F_{\m\n}F^{\m\n}+\theta^{\m\n}\,\Tr\int d^4x\,
\Big(\frac 14 F_{\m\n}F_{\r\s}-
F_{\m\r}F_{\n\s} \Big)F^{\r\s},
\label{act}
\end{equation}
which is  the starting point for the analysis of  $\theta$-expanded
noncommutative gauge models.
The action consists of two terms. The first term is the ordinary commutative 
action, and the second gives additional interactions which describe
noncommutativity in  the leading order in $\theta$.
In order to take into account the nonuniqueness of the expansions
(\ref{expansion}-\ref{FF}),
one should also add  terms which correspond to the freedom
 (\ref{nonuniq}). In the action this amounts to
\begin{equation}
S^\prime = S -\Tr\int d^4x F^{\mu\nu}D_\mu {X_\nu}.
\end{equation}

The additional terms which could be included in the Lagrangian (\ref{act}), that 
is those linear in $\theta$ and of correct dimension are,
\begin{equation}
F^{\mu\nu} D_\mu {X_\nu}= F^{\mu\nu} D_\mu\left(  b_1\,\theta^{\rho\sigma}D_\nu 
F_{\rho\s}+\,b_2\, \theta^\rho{}_\nu D^\sigma F_{\rho\s}
+\,b_3\, \theta^{\rho\sigma}D_\rho F_{\n\s} \right) .
\label{X}
\end{equation}
Out of these three terms  the second  vanishes owing to
its symmetry-antisymmetry properties.
The third term can be transformed into the first one
using the Bianchi identities
\footnote{One could in principle also add the parity violating
terms. There are two independent expressions:
$ F^{\mu\nu} D_\mu \epsilon^{\rho\s\a\b}\theta_{\a\b}D_\n F_{\r\s}$ and
$ F^{\mu\nu} D_\mu \epsilon^{\n\s\a\b}\theta_{\r\b}D^\r F_{\a\s}$.
These terms violate parity if one assumes that $\theta^{\m\n}$  is
invariant under parity; compare, however, with \cite{Aschieri:2002mc}.
We shall not discuss such a possibility in this article.}.

 In summary, the freedom due to the SW field redefinitions reduces
to the possibility to add one term, $\Delta S$,  to the original Lagrangian:
\begin{equation}
\Delta S = -2b\,\theta^{\rho\sigma}\,\Tr\int d^4x\, F^{\mu\nu}D_\mu D_\nu 
F_{\rho\s}= b \,\theta^{\rho\sigma}\,\Tr\int d^4x \,F^{\mu\nu}F_{\m\n} 
F_{\rho\s}.\end{equation}
Writing $b = -\frac 14 +\frac a4$, we obtain the following general form of
the noncommutative gauge field action:
\begin{equation}
S =-\frac 12\Tr\int d^4x\,F_{\m\n}F^{\m\n}+\theta^{\m\n}\,\Tr\int d^4x\,
\big(\frac a4 \, F_{\m\n}F_{\r\s}-
F_{\m\r}F_{\n\s} \big)F^{\r\s} .
\label{Act}
\end{equation}
The coefficient $a$ is going to be fixed by the requirement of
renormalizability in the next section.

\subsection{$\rm U(1)_Y\otimes SU(2)_L\otimes SU(3)_C$}

The discussion given above was a general one, without any specification of
the gauge group $G$ or of its representations. However, as the $\theta$-linear 
term in the action
includes the trace of the product of three group generators,  it is obvious that
the action  is  a representation-dependent quantity.
In the commutative case, the action contains only
the trace of  the product of two generators which is up to normalization
the same for all group representations,
$\Tr T^a T^b \sim \delta ^{ab}$ (if we assume
the usual properties of $G$, i.e. that it is semisimple, compact, etc.).
But in (\ref{Act}) we    have a factor $\Tr \{T^a,T^b\}T^c \sim d^{abc}$.
One could perhaps assume that, as the field strength transforms according
to the adjoint representation,
the symmetric coefficients $d^{abc}$ are given in that representation.
However, when the matter fields are included, other representations
of $G$ are present too, and
therefore the expression (\ref{Act}) is ambiguous.

To start the discussion of the gauge field action-dependence
on the gauge group and/or on its representation, we use
the most general form of the action, \cite{Aschieri:2002mc}:
\be
S_{cl} = -\frac{1}{2}  \int
d^4x \, \sum_{\car} {C_{\car}} {\Tr}\Big(
{\car}(\widehat F_{\mu \nu}) * {\car}(\widehat F^{\mu
\nu})\Big).
\label{action1}
\ee
The sum is, in principle, taken  over all irreducible representations
$\car$  of $G$  with arbitrary weights $C_\car$.
Of course, for the gauge group $G$ we take $\rm U(1)_Y\otimes SU(2)_L\otimes 
SU(3)_C$. To relate the action (\ref{Act}) to the usual action of the 
commutative standard model, we make the decompositions
\bea
V_\m &=& g^\prime \ca _\m \car (Y) + g B_\m^i \car (T^i_L) +g_S G_{\m}^a\car 
(T^a_S),\\
F_{\mu\nu} &=&g^\prime f_{\mu\nu}\car(Y)+g B_{\m\n}^i\car (T^i_L)+g_S 
G_{\m\n}^a\car(T^a_S).\eea
The $\car(Y)$, $\car (T_L^i)$,  $\car (T_S^a)$ denote the representations of
the group generators $Y$, $T_L^i$ and $T_S^a$ of $\rm U(1)_Y$, $\rm SU(2)_L$
and $\rm SU(3)_C$, respectively; the group indices run as
$i,j = 1,\dots 3$ and  $a,b = 1,\dots 8$.
According to \cite{Aschieri:2002mc}, we take that  $C_\car $ are nonzero only
for the particle representations which are present in the standard model.
Then from (\ref{action1}) we obtain the expression
for the  $\theta$-independent part of the Lagrangian
\begin{eqnarray}
\cl_{SM}
&=&-\frac12g^{\prime 2}\sum_{\car}C_{\car}
d(\car_2)d(\car_3)\car_1(Y)\car_1(Y)\,f_{\m\n}
f^{\m\n}\nonumber \\
&-\frac12&g^2\sum_{\car}C_{\car}d(\car_3)\Tr(\car(T_L^i)\car(T_L^j))\,
B_{\m\n}^i
B^{\m\n j}\nonumber\\
&-&\frac12g^2_S\sum_{\car}C_{\car}d(\car_2)\Tr(\car(T_S^a)\car(T_S^b))\,
G_{\m\n}^a
G^{\m\n b},
\label{LSM}
\end{eqnarray}
where $d(\car)$ denotes the dimension of the representation $\car$.
Identifying (\ref{LSM}) with the SM Lagrangian,
we find that  the weights have to be constrained to match  the coupling
constants in the standard model  in the following way
\cite{Calmet:2001na,Aschieri:2002mc,Goran}:
\bea
\frac{1}{2g^{\prime 2}} &=& \sum_{\car}C_{\car} \label{match1}
d(\car_2)d(\car_3)\car_1(Y)^2,
\\
\frac{1}{g^2}\frac{\delta^{ij}}{2} &=& \sum_{\car}C_{\car}
d(\car_3)\Tr(\car(T_L^i)\car(T_L^j)), \label{match2}
\\
\frac{1}{g^{ 2}_S}\frac{ \delta^{ab}}{2} &=& \sum_{\car}C_{\car}
d(\car_2)\Tr(\car(T_S^a)\car(T_S^b)). \label{match3}
\eea

The noncommutative correction, that is the $\theta$-linear part of the 
Lagrangian, reads\bea
\cl^\theta &=& \sum \cl^\theta _i =g^{\prime 3}\kappa_1\theta^{\m\n}
\left( \frac a4 f_{\m\n}f_{\r\s}f^{\r\s}-f_{\m\r}f_{\n\s}f^{\r\s}\right)
\nonumber\\
&+&g^3 \kappa_4^{ijk}\theta^{\m\n}
 \left(\frac a4 B_{\m\n}^i B_{\r\s }^j B^{\r\s k} - B_{\m\r}^i B_{\n\s }^j 
B^{\r\s k} \right) \nonumber\\
&+& g^3_S\kappa_5^{abc}\theta^{\m\n}
 \left( \frac{ a}{4}G_{\m\n}^a G_{\r\s }^b G^{\r\s c}- G_{\m\r}^a G_{\n\s }^b 
G^{\r\s c}\right) \nonumber \\
&+& g^\prime g^2  \kappa_2 \theta^{\m\n}
\left(\frac a4 f_{\m\n}B_{\r\s }^i B^{\r\s i} -f_{\m\r}B_{\n\s }^i B^{\r\s i} + 
c.p.\right) \nonumber \\
&+& g^\prime g^2_S  \kappa_3 \theta^{\m\n}
\left(\frac a4 f_{\m\n}G_{\r\s }^a G^{\r\s a} -f_{\m\r}G_{\n\s }^a G^{\r\s a} 
+c.p. \right),\label{lterm}
\eea
%%%%%%%%%%%%%%%%%%%
\TABULAR{|c|c|c|c|c|c|}
{\hline  %\addtolength{\myVSpace}{1mm}
\xstrut
  & $\rm SU(3)_C$ & $\rm SU(2)_L$ & $\rm U(1)_Y$  & $\rm U(1)_Q$
  & $\rm T_3$
   \\
\hline \xstrut
     $ e_R$
   & ${\bf 1}$
   & ${\bf 1}$
   & $-1$
   & $-1$
   & $0$
   \\[6pt]
   $ L_L=\left(\begin{array}{c} \nu_L \\ e_L \end{array} \right )$
   & ${\bf 1}$
   & ${\bf 2}$
   & $-1/2$
   & $\left(\begin{array}{c} 0 \\ -1 \end{array} \right )$
   & $\left(\begin{array}{c} 1/2 \\ -1/2 \end{array} \right )$
    \\ [20pt]
   $u_R$
   & ${\bf 3}$
   & ${\bf 1}$
   & $2/3$
   & $2/3$
   & $0$
   \\[10pt]
  %\hline
   $d_R$
   & ${\bf 3}$
   & ${\bf 1}$
   & $-1/3$
   & $-1/3$
   & $0$
    \\[10pt]
     $Q_L=\left(\begin{array}{c}  u_L \\ d_L \end{array} \right )$
   & ${\bf 3}$
   & ${\bf 2}$
   & $1/6$
   & $\left(\begin{array}{c} 2/3 \\ -1/3 \end{array} \right )$
   & $\left(\begin{array}{c} 1/2 \\ -1/2 \end{array} \right )$
    \\ [20pt]
$\Phi=\left(\begin{array}{c}  \phi^+ \\  \phi^0 \end{array} \right )$
   & ${\bf 1}$
   & ${\bf 2}$
   & $1/2$
   & $\left(\begin{array}{c} 1 \\ 0 \end{array} \right )$
   & $\left(\begin{array}{c} 1/2 \\ -1/2 \end{array} \right )$
 \\[20pt]  \hline }
{Matter fields of the first generation.
 Electric charge is given by the Gell-Mann-Nishijima relation
 $ Q=T_3+Y$.}
%%%%%%%%%%%%%%%%%%%
where the $c.p.$ in (\ref{lterm}) denotes the addition of the terms
obtained by a cyclic permutation of
fields without changing the positions of indices.
The couplings in (\ref{lterm}) are defined as follows:
\bea
\kappa_1 &=&\sum_{\car}C_{\car}
d(\car_2)d(\car_3)\car_1(Y)^3,
\label{k1} \\
\kappa_2 \delta^{ij} &=& \sum_{\car}C_{\car}
d(\car_3)\car_1 (Y)\Tr (\car_2(T^i_L)\car_2(T^j_L)),
\label{k2} \\
\kappa_3 \delta^{ab}&=& \sum_{\car}C_{\car}
d(\car_2)\car_1 (Y)\Tr (\car_3(T^a_S)\car_3(T^b_S)),
\label{k3} \\
\kappa_4 ^{ijk}&=& \frac12\sum_{\car}C_{\car}
d(\car_3)\Tr (\{\car_2(T^i_L),\car_2(T^j_L)\}\car_2(T^k_L)),
\label{k4} \\
\kappa_5^{abc} &=&\frac12 \sum_{\car}C_{\car}
d(\car_2)\Tr (\{\car_3(T^a_S),\car_3(T^b_S)\}\car_3(T^c_S)) .
\label{k5}
\eea

Let us  discuss the dependence of $\kappa_1, \dots  ,\kappa_5$ on the 
representations of matter fields.
For the first generation of the standard model there are six such 
representations,summarized in Table 1; they  produce six
independent constants $C_\car$\footnote{We assume that $C_\car >0$;
therefore the six $C_\car$'s were denoted by
 $\frac{1}{g_i^2}\,,i=1,...,6$, in \cite{Calmet:2001na,Goran}.}.
These constants  are already constrained by the three relations
(\ref{match1}-\ref{match3}). The couplings $\kappa_1,\dots ,
\kappa_5$ given by (\ref{k1}-\ref{k5}) also depend on $C_\car$. However, one can
immediately verify that $\kappa_4^{ijk} = 0$.
This follows from the fact that the symmetric coefficients
$d^{ijk}$ of $\rm SU(2)$ vanish for all irreducible representations. We shall in 
addition take that $\kappa^{abc}_5 = 0$. The argument for this assumption is 
related to the invariance of thecolour sector of the SM  under charge 
conjugation.Although apparently in Table 1 one has only the fundamental
 representation {\bf 3} of $\rm SU(3)_C$, there are in fact both ${\bf 3}$ and 
${\bf \bar 3}$ representations with the same weights, $C_{{\bf 3}} =C_{{\bf\bar 
3}}$. In the Lagrangian this corresponds to  writing each minimally-coupled 
quark termas a half of the sum of the original  and the charge-conjugated terms.
Since the symmetric coefficients for the {\bf 3} and ${\bf \bar 3}$  
representations satisfy$d^{abc}_{{\bf \bar 3}} = - d^{abc}_{{\bf 3}}$,
we obtain
\begin{equation}
 \kappa_5^{abc} = C_{{\bf 3}}d_{\bf 3}^{abc} + C_{{\bf \bar 3}}d_{\bf \bar 
3}^{abc} =0 .\label{Kappa}
\end{equation}

We are left only with three nonvanishing couplings, $\kappa_1$, $\kappa_2$
and $\kappa_3$, depending on six  constants $C_1,\dots ,C_6$ (indices $1,\dots 
,6$ enumerate the representations  as they are given in Table 
1):\bea
\kappa_1&=&-\,C_1-\frac14\,C_2+\frac89 \,C_3-\frac19 \,C_4+
\frac{1}{36} \,C_5+\frac14 \,C_6\,,
\nonumber \\
\kappa_2&=&-\,\frac14\,C_2+\frac14 \,C_5+\frac14 \,C_6\,,
\nonumber \\
\kappa_3&=& +\,\frac13 \,C_3- \frac16\,C_4+ \frac16 \,C_5\,.
\label{kappa}
\eea
There are three relations among $C_i$'s:
\bea
\frac{1}{g^{\prime 2}}&=&\,2\,C_1+C_2+\frac83 \,C_3+\frac23 \,C_4+
\frac13 C_5+C_6\,,
\nonumber \\
\frac{1}{g^{2}}&=&\,C_2+3\,C_5+\,C_6\,,
\nonumber \\
\frac{1}{g^2_s}&=& \,C_3+ \,C_4+2 \,C_5\,,
\label{C}
\eea
in effect representing three consistency conditions imposed on
(\ref{Act}) in a way to match the SM action at zeroth order in $\theta$.
Note that detailed discussions about the solutions of the system of three 
equations (\ref{kappa}) and six unequations $C_i >0$, satisfying (\ref{C}), are 
given in \cite{Goran}.
 Our classical noncommutative action reads
\be
S_{cl} = S_{SM} +S^\theta ,
\label{Scl}
\ee with
\bea S^\theta &=& \sum_{i=1}^3
S^{\theta}_i = g^{\prime
3}\kappa_1\theta^{\m\n} \int d^4 x\left( \frac a4
f_{\m\n}f_{\r\s}f^{\r\s}-f_{\m\r}f_{\n\s}f^{\r\s}\right)\nonumber\\
&+& g^\prime
g^2  \kappa_2 \theta^{\m\n}\int d^4 x\left(\frac a4 f_{\m\n}B_{\r\s }^i B^{\r\s
i} -f_{\m\r}B_{\n\s }^i B^{\r\s i} + c.p.\right) \nonumber \\[4pt]
&+& g^\prime g^2_S
\kappa_3 \theta^{\m\n}\int d^4 x\left(\frac a4 f_{\m\n}G_{\r\s }^a G^{\r\s a}
-f_{\m\r}G_{\n\s }^a G^{\r\s a} +c.p. \right).
\label{St}
\eea
\smallskip

\EPSFIGURE{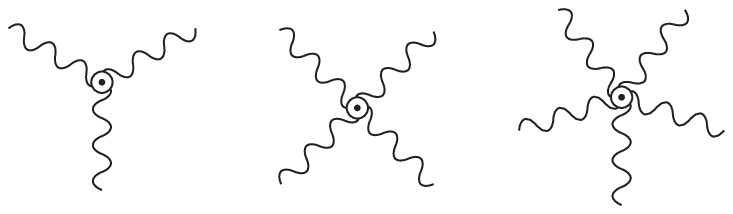}{$\theta$-vertices}
The noncommutative couplings introduce additional vertices, as depicted in 
Figure 1. For simplicity, we do not distinguish the
gauge fields $\ca_\mu$,  $B_\mu^i$ and $G_\mu^a$
by different types of lines: the dependence on the
fields is  not difficult to trace.

The term $S^{\theta}_1$ in (\ref{St})  is one-loop renormalizable to
linear order in $\theta$ \cite{Buric:2005xe} since the one-loop correction
to the $S^{\theta}_1$ is of the second order in $\theta$.
We need to investigate only the
renormalizability of remaining $S^{\theta}_2$ and $S^{\theta }_3$ parts of
the action (\ref{St}).

\section{One-loop renormalizability}

\subsection{Effective action}

We compute the divergencies in the one-loop effective action using
the background-field method \cite{'tHooft:1973us,PS}. As we have
already explained many details of similar calculations
\cite{Maja},  here we just introduce  the notation. Let the
classical action  be given by $S_{cl}[\phi]$;  in our case, the
fields are, $\phi_A = (\ca_\m , \, B_\m^i, \,  G_\m^a \, ) $. To
quantize, one performs the functional integral. The integral over
the quantum fields, ${\bf \Phi}_A$, can be calculated in the
saddle-point approximation around the classical (background)
configuration, denoted also by $\phi _A$. The effective action is
\begin{equation}
\Gamma [\phi] =
S_{cl}[\phi] + \Gamma^{(1)} [\phi].
\label{GS2}
\end{equation}
The first quantum correction to the one-loop effective
action $\Gamma^{(1)} [\phi]$, is given by
\begin{equation}
\Gamma ^{(1)}[\phi] =\frac{i}{2}\log\det
S^{(2)}_{cl}[\phi]=\frac{i}{2}\Tr\log S^{(2)}_{cl}[\phi].
\label{gama1}
\end{equation}
In (\ref{gama1}) the $ S^{(2)}_{cl}[\phi]$
is the second functional derivative of the classical action,
\be
S^{(2)}_{cl}[\phi]= \frac {\delta^2 S_{cl}}{\delta\phi_A \delta\phi _B} .
\label{S2}
\ee

In the case of the polynomial interactions as we have in (\ref{St}),
one can find $S^{(2)}_{cl}$  simply by splitting the fields
into the classical-background  plus the quantum-fluctuation parts, that is,
$\phi_A\to \phi_A + {\bf\Phi}_A$, and by computing the terms quadratic in the 
quantum fields. For the action (\ref{Act}), the classical Lagrangian reads
\bea
\cl_{cl} &=& \cl_{SM} +\sum \cl^\theta_i
\nonumber\\
&=&
-\frac 14 f_{\m\n}f^{\m\n} - \frac 14 B_{\m\n}^i B^{\m\n i} -\frac 14G_{\m\n}^a 
G^{\m\n a}\nonumber\\
&+& g^{\prime 3}  \kappa_1 \theta^{\m\n}
\left(\frac a4 f_{\m\n}f_{\r\s } f^{\r\s } -f_{\m\r}f_{\n\s } f^{\r\s } \right)
\nonumber\\
&+& g^\prime g^2  \kappa_2 \theta^{\m\n}
\left(\frac a4 f_{\m\n}B_{\r\s }^i B^{\r\s i} -f_{\m\r}B_{\n\s }^i B^{\r\s i} + 
c.p.\right)\nonumber\\
&+& g^\prime g_S^2  \kappa_3 \theta^{\m\n}
\left(\frac a4 f_{\m\n}G_{\r\s }^a G^{\r\s a} -f_{\m\r}G_{\n\s }^a G^{\r\s a} + 
c.p.\right).\label{claction}
\eea
Writing the $c.p.$ terms in (\ref{claction}) explicitly, we obtain
\bea
\cl_{cl}
&=&
-\frac 14 f_{\m\n}f^{\m\n} - \frac 14 B_{\m\n}^i B^{\m\n i} -\frac 14G_{\m\n}^a 
G^{\m\n a}\\
&+& g^{\prime 3}  \kappa_1 \theta^{\m\n}
\left(\frac a4 f_{\m\n}f_{\r\s } f^{\r\s } -f_{\m\r}f_{\n\s } f^{\r\s } \right)
\nonumber\\
&+& g^\prime g^2  \kappa_2 \theta^{\m\n}
\big(\frac a4 f_{\m\n}B_{\r\s }^i B^{\r\s i} - 2f_{\m\r}B_{\n\s }^i B^{\r\s i}
+ \frac a2 f_{\r\s}B_{\m\n }^i B^{\r\s i} - f_{\r\s}B_{\m\r }^i B^{\n\s i}
 \big)
\nonumber\\
&+& g^\prime g^2_S  \kappa_3 \theta^{\m\n}
\big(\frac a4 f_{\m\n}G_{\r\s }^a G^{\r\s a} - 2f_{\m\r}G_{\n\s }^a G^{\r\s a}
+ \frac a2 f_{\r\s}G_{\m\n }^a G^{\r\s a} - f_{\r\s}G_{\m\r }^a G^{\n\s a}
 \big) \,,
 \nonumber
 \label{lagr}
\eea
the classical Lagrangian which we are using next in the renormalization 
procedure.
\subsection{Interaction vertices}

In order to fix the quantum gauge symmetry, we have to add the
gauge-fixing term to the Lagrangian (\ref{lagr}). The
gauge-fixing term is added to the $\theta$-independent part in the
usual way, \cite{PS, Maja}. After making the splitting \be
\ca_\m\to \ca_\m +{\bf A}_\mu,\ \ B_\m^i\to B_\m^i +{\bf B}^i_\m,\
\ G_\m^a\to G_\m^a +{\bf G}^a_\m, \ee we obtain for the  quadratic
part  of the  action (\ref{lagr}): \be \frac12\left( \matrix{{\bf
A}_\a\,{\bf B}_\a^i\, {\bf G}_\a^a}\right)
\left(\matrix{&g^{\a\b}\Box+M^{\a\b}&*&*\cr
&*&g^{\a\b}\d^{ij}\Box+V^{\a\b;ij}&0\cr&*&
0&g^{\a\b}\d^{ab}\Box+W^{\a\b ab}\cr}\right)\left( \matrix{{\bf
A}_\b\cr {\bf B}_\b^j\cr {\bf G}_\b^b}\right). \label{matrix} \ee
In (\ref{matrix}), $*$ stands for the terms which will not
contribute to linear order: they give higher-order corrections.
The first matrix element in (\ref{matrix}) is given by $M^{\a\b}=
\overleftarrow { \pa _\m} M^{\m\a,\n\b}(x)\overrightarrow { \pa
_\n}$, where \bea M^{\m\r,\n\s}&=& {1\over 2}\,
(g^{\m\n}g^{\r\s}-g^{\m\s}g^{\n\r})\theta^{\a\b}f_{\a\b}\ncr
&+&g^{\m\n}(\theta^{\a\r}{f^\s}_\a+\theta^{\a\s}{f^\r}_\a)
+g^{\r\s}(\theta^{\a\m}{f^\n} _\a+\theta^{\a\n}{f^\m} _\a )\ncr
&-&g^{\m\s}(\theta^{\a\r}{f^\n}_\a+\theta^{\a\n}{f^\r}_\a)
-g^{\n\r}(\theta^{\a\s}{f^\m} _\a+\theta^{\a\m}{f^\s} _\a )\ncr
&+& \theta^{\m\r}f^{\n\s}+\theta^{\n\s}f^{\m\r}-
\theta^{\r\s}f^{\m\n} -\theta^{\m\n}f^{\r\s}
-\theta^{\n\r}f^{\m\s} -\theta^{\m\s}f^{\n\r} \ .\label{M}\eea The
structure of $V^{\a\b;ij} $ is as follows: \be
V^{\a\b;ij}=(N_1+N_2+T_1+T_2+T_3)^{\a\b;ij}. \label{V} \ee The
operators $N_1$ and $N_2$ come from the commutative 3-vertex and
4-vertex interactions: \bea (N_1)^{
ij}_{\a\b}&=&-2ig_{\a\b}(B_\m)^{ij}\pa^\m-i(\pa^\m
B_\m)^{ij}g_{\a\b},
\label{N1}\\
(N_2)^{ ij}_{\a\b}&=&-(B_\m B^\m)^{ij}g_{\a\b}-2i(B_{\a\b})^{ij}\,,
\label{N2}
\eea
where we have used the notation $(X_{\mu})^{ij}=-if^{ijk}X^k_{\mu}$.
The operators $T_1$, $T_2$ and $T_3$ describe the  $\theta$-linear,
that is the  noncommutative vertices.
They are more involved:
\bea
(T_1)_{\a\b}^{ij}&=&g^\prime g^2  \kappa_2
\d^{ij}\Big[a(
\overleftarrow { \pa _\m}\theta^{\r\s}f_{\r\s}g_{\a\b}\overrightarrow { \pa
_\m}-\overleftarrow { \pa _\b}\theta^{\r\s}f_{\r\s}\overrightarrow { \pa
_\a})
\label{T1}\\
&-&
2(\overleftarrow { \pa _\b}\theta_{\r \a} f^{\m
\r}\overrightarrow { \pa _\m}-\overleftarrow { \pa ^\n}\theta^{\r}_{\ \a} f_{\b
\r}\overrightarrow { \pa _\n}-\overleftarrow { \pa _\s}\theta^{\r\s}f_{\m
\r}g_{\a\b}\overrightarrow { \pa ^\m}
+
\overleftarrow { \pa
_\s}\theta^{\r\s}f_{\b \r}\overrightarrow { \pa _\a}
\nonumber\\
&+&
\overleftarrow { \pa _\m}\theta_{\r \b} f^{\m
\r}\overrightarrow { \pa _\a}-\overleftarrow { \pa ^\n}\theta^{\r}_{\ \b} f_{\a
\r}\overrightarrow { \pa _\n}-\overleftarrow { \pa ^\m}\theta^{\r\s}f_{\m
\r}g_{\a\b}\overrightarrow { \pa _\s}
\nonumber\\
&+&
\overleftarrow { \pa_\b}\theta^{\r\s}f_{\a \r}\overrightarrow { \pa _\s}) +
2a(\overleftarrow { \pa _\r}\theta^{\r}_{\ \a} f_{\m \b}\overrightarrow { \pa
^\m}+\overleftarrow { \pa ^\m}\theta^{\r}_{\ \b} f_{\m \a}\overrightarrow { \pa
_\r})
\nonumber\\
&-&
2(\overleftarrow { \pa_\m}\theta_{\a\b}f^{\m \n}\overrightarrow { \pa 
_\n}-\overleftarrow{\pa^\m}\theta_{\a\s}f_{\m\b} \overrightarrow { \pa ^\s} - 
\overleftarrow { \pa^\s}\theta_{\b\s}f_{\m \a}\overrightarrow { \pa 
^\m}+\overleftarrow { \pa_\r}\theta^{\r\s}f_{\a\b }\overrightarrow { \pa 
_\s})\Big],\nonumber\\
(T_2)^{ij}_{\a\b}&=&g^\prime g^2i  \kappa_2\Big[a(-\overleftarrow { \pa
_\m}\theta^{\r\s}g_{\a\b}f_{\r\s }(B^\m)^{ij}-\theta^{\r\s}f_{\r\s
}g_{\a\b}(B^\m)^{ji}\overrightarrow { \pa _\m}
\label{T2}\\
&+&
\overleftarrow { \pa_\b}\theta^{\r\s}f_{\r\s }(B_\a)^{ ij}+\theta^{\r\s}f_{\r\s
}(B_\b)^{ji}\overrightarrow { \pa _\a}+\theta_{\r\s}f^{\r\s}(B_{\a\b})^{ij})
\nonumber\\
&-&
2( -\overleftarrow { \pa _\b}\theta_{\r\a}f^{\m\r }(B_\m)^{ij}
-\theta_{\r\b}f^{\m\r }(B_\m)^{ ji}\overrightarrow { \pa_\a}
+\overleftarrow { \pa _\n}\theta_{\r\a}f_{\b}^{\ \r}(B^\n)^{ij}
\nonumber\\
&+&
%\overleftarrow { \pa _\n}\theta_{\r\a}f_{\b}^{\ \r}(B^\n)^{ij}+
\theta_{\r\b}f_{\a}^{\ \r }(B^\n)^{ji}\overrightarrow { \pa
_\n}+\overleftarrow { \pa _\s}\theta^{\r\s}f_{\m\r
}g_{\a\b}(B^\m)^{ij}+\theta^{\r\s}f_{\m\r }g_{\a\b}(B^\m)^{ji}
\overrightarrow {\pa _\s}
\nonumber\\
&-&
\overleftarrow { \pa _\s}\theta^{\r\s}f_{\b\r }(B_\a)^{ij}
-\theta^{\r\s}f_{\a\r }(B_\b)^{ ji}\overrightarrow { \pa _\s}-\overleftarrow
{ \pa _\m}\theta_{\r\b}f^{\m\r }(B_\a)^{ ij}-\theta_{\r\a}f^{\m\r
}(B_\b)^{ ji}\overrightarrow { \pa _\m}
\nonumber\\
&+&
\overleftarrow { \pa_\m}\theta^{\r\s}g_{\a\b}f^\m_{\ \r 
}(B_\s)^{ij}+\theta^{\r\s}f_{\m\r}g_{\a\b}(B_\s)^{ji}\overrightarrow { \pa 
^\m}+\overleftarrow { \pa_\m}\theta^{\r}_{\ \b}f_{\a\r }(B^\m)^{ij}
%+\theta_{\r\a}f_{\b}^{\ \r}
%(B^\m)^{ji}\overrightarrow { \pa _\m}
\nonumber\\
&+&
\theta_{\r\a}f_{\b}^{\ \r}
(B^\m)^{ji}\overrightarrow { \pa _\m}
-\overleftarrow { \pa_\b}\theta^{\r\s}f_{\a\r
}(B_\s)^{ij}-\theta^{\r\s}f_{\b\r}(B_\s)^{ji}\overrightarrow { \pa
_\a}+\theta^{\r\s}f_{\a\r}(B_{\b\s})^{ij}
\nonumber\\
&+&
\theta_{\r\b}f^{\m\r}(B_{\m \a})^{ij}
+ \theta^{\r\s}f_{\b\r}(B_{\a\s})^{ji}
\nonumber\\
&+&
\theta_{\r\a}f^{\m\r}(B^\m_{\ \b})^{ji})-2a(\overleftarrow { \pa
^\r}\theta_{\r\a}f_{\m\b}(B^\m)^{ij}+\theta_{\r\b}f_{\m\a
}(B^\m)^{ji}\overrightarrow { \pa ^\r}
\nonumber\\
&+&
\overleftarrow { \pa^\m}\theta_{\r\b}f_{\m\a
}(B^\r)^{ij}+\theta_{\r\a}f_{\m\b}(B^\r)^{ji}\overrightarrow { \pa^\m}
-\frac12\theta_{\r\s}f_{\a\b}(B^{\r\s})^{ij}
\nonumber\\
%-\frac12\theta_{\r\s}f_{\a\b}(B^{\r\s})^{ij}
&-&
\frac12\theta_{\a\b}f_{\r\s}(B^{
\r\s})^{ij})-2(-\overleftarrow { \pa ^\m}\theta_{\a\b}f_{\m\n
}(B^\n)^{ij}-\theta_{\b\a}f_{\m\n }(B^\n)^{ji}\overrightarrow { \pa
^\m}
\nonumber \\
&+&
\overleftarrow { \pa ^\m}\theta_{\a\s}f_{\m\b
}(B^\s)^{ij}+\theta_{\b\s}f_{\m\a }(B^\s)^{ji}\overrightarrow { \pa
^\m}+\overleftarrow { \pa ^\r}\theta_{\r\b}f_{\a\n
}(B^\n)^{ij}+\theta_{\r\a}f_{\b\n }(B^\n)^{ji}\overrightarrow { \pa
^\r}
\nonumber\\
&-&
\overleftarrow { \pa _\r}\theta_{\b\s}f_{\a\b }(B_\s)^{ij}
-\theta^{\r\s}f_{\b\a}(B_\s)^{ji}\overrightarrow { \pa
_\r}+\theta_{\b\s}f_{\a\n}(B^{\n\s})^{ij}+
\theta_{\a\s}f_{\b\n}(B^\n_{\ \s})^{ji})
\Big],
\nonumber\\
(T_3)_{\a\b}^{ ij}
&=&g^\prime g^2  \kappa_2\Big[a(\theta^{\r\s}f_{\r\s}(B_\m
B^\m)^{ij}g_{\a\b}-\theta^{\r\s}f_{\r\s}(B_\b
B_\a)^{ij})
\label{T3}\\
&-&
2(\theta_{\r\a}f^{\m\r}(B_\b B_\m)^{ij}-\theta^{\r}_{\
\a}f_{\b\r}(B_\n B^\n)^{ij}-\theta^{\r\s}f_{\m\r}(B_\s
B^\m)^{ij}g_{\a\b}
\nonumber\\
&+&
\theta^{\r\s}f_{\b\r}(B_\s B_\a)^{ij}+(\a
\leftrightarrow\b\ \  i\leftrightarrow j))
\nonumber\\
&+&
2a(\theta_{\r\a}f_{\m\b}(B^\r
B^\m)^{ij}+2\theta_{\r\b}f_{\m\a}(B^\r
B^\m)^{ji})
\nonumber\\
&-&2(\theta_{\a\b}f^{\m\n}(B_\m
B_\n)^{ij}-\theta_{\a\s}f_{\m\b}(B^\m B^\s)^{ij}
-\theta_{\b\s}f_{\m\a}(B^\m B^\s)^{ji}+\theta^{\r\s}f_{\a\b}(B_\r
B_\s)^{ij})\Big].
\nonumber
\eea

We do not write the matrix $W^{\a\b, ab}$ explicitly
as it is completely analogous to $V^{\a\b, ij}$
up to the change $B_\m^i\leftrightarrow G_\m^a$.

\subsection{Divergencies}

We compute the divergencies due to the $\,\rm U(1)_Y-SU(2)_L\,$ part of
the noncommutative action, $S^\theta_2$.
The result for $\,\rm U(1)_Y-SU(3)_C\,$ is analogous and follows immediately. 
The one-loop effective action  is
\bea
\G^{(1)}_{\theta,2}&=&
\frac i2 \Tr \log \left(\ci + \Box ^{-1} (N_1+N_2+T_1+T_2+T_3)\right)
\\
&=&\frac i2\, \sum_{n=1}^\infty \frac{(-1)^{n+1}}{n}
\Tr \left( \Box ^{-1}N_1+ \Box ^{-1} N_2+\Box ^{-1}T_1+ \Box ^{-1}T_2+ \Box 
^{-1}T_3 \right)^n .\label{trlog} \nonumber
\eea
For dimensional reasons, the divergencies in  $\theta$-linear
order are all of the form $\theta f B^2$. Consequently, from the  sum 
(\ref{trlog}) we need to  extract and compute only terms that  contain three 
external fields. A careful analysis gives that these terms are
\be
\G^{(1)}_{\theta,2}=\frac{i}{2}\Tr [(\Box^{-1}N_1)^2\Box^{-1} 
T_1-\Box^{-1}N_1\Box^{-1}T_2-\Box^{-1}N_2\Box^{-1} T_1].             \label{333}
\ee
As one can readily see, only the vertices obtain divergent contributions.
For the $\theta$-3-vertex, the diagrams
which correspond to the traces in (\ref{trlog}) are given in Figure 2.
\EPSFIGURE{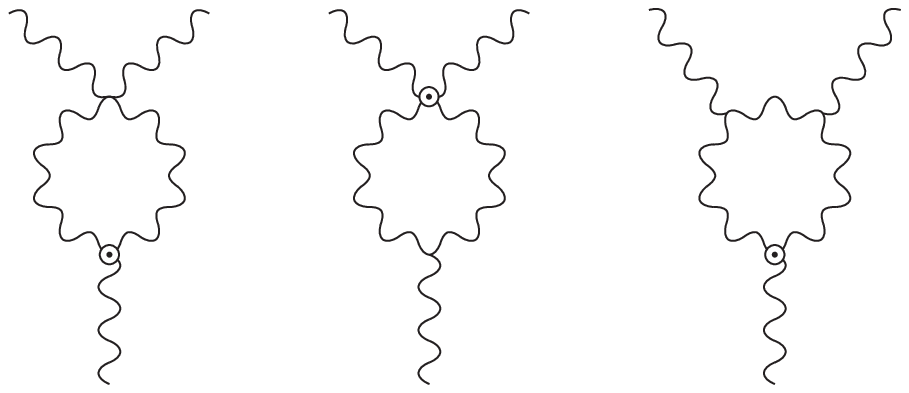}{One-loop divergent corrections to the $\theta$-3-vertex.}
Being written in terms of the field strengths, that is covariantly, 
(\ref{trlog}) also contains the contributions to the $\theta$-4-vertex  and 
$\theta$-5-vertex.We do not draw the corresponding diagrams: they can be easily 
obtained from Figure 2 by adding external legs (in accordance with the Feynman 
rules).
The divergent part of (\ref{333}) is calculated in
the momentum representation by dimensional regularization. The results are given 
by\bea
\Tr(\Box^{-1}N_1\Box^{-1}T_2)
&=&
\frac{4i}{3(4\pi)^2\epsilon}g^\prime g^2
\kappa_2
\nonumber\\
&\times&
\Big[(6-2a)(\theta^{\r\s}f_{\a\r}+\theta_{\r\a}f^{\s\r})(B^{\a
i}\pa_\m\pa_\s B^{\m i}-B^{\a i}\Box B^{ i}_\s)
\nonumber\\
&+&
(3a-4)\theta^{\r\s}f_{\r\s}(B^{\n i}\pa_\m\pa_\n B^{\m
i}-B_{\m}^{i}\Box B^{\m i})\Big],
\label{N1T2}\\[10pt]
\Tr(\Box^{-1}N_2\Box^{-1} T_1)
&=&
\frac{4i}{3(4\pi)^2\epsilon}g^\prime g^2
\kappa_2
\nonumber\\
&\times&
\Big[(2a-6)(\theta^{\r\s}f_{\a\r}
+\theta_{\r\a}f^{\s\r})(B^{\n i}\pa_\s\pa^\a B_\n^i
+\pa_\s B^{\m i}\pa^\a B_{\m}^i)
\nonumber\\
&+&
\theta^{\r\s}f_{\r\s}(18-11a)(\pa_\n B^{\n i}\pa_\m
B^{\m i}+B_{\m }^i\Box B^{\m i})\Big] ,
\label{N2T1}\\[10pt]
\Tr(\Box^{-1}N_1^2\Box^{-1} T_1)
&=&
\frac{4i}{3(4\pi)^2\epsilon}
g^\prime g^2\kappa_2\Big[\theta^{\r\s}f_{\r\s}\Big((22-14a)B_{\m}^i\Box B^{\m i}
\nonumber\\
&+&
(15-10a)\pa_\n B^{\m i}\pa^\n  B_{ \m}^i
\nonumber\\
&+&
(3a-4)B^{\m i}\pa_\m \pa_\n B^{\n i}+(3-a)\pa_\m B^{\n  i}\pa_\n  B^{\m i}\Big)
\nonumber\\
&+&
(\theta^{\r\s}f_{\a\r}+\theta_{\r\a}f^{\s\r})\Big((2a-6)(B_{\s}^i\Box
B^{\a i}\nonumber\\&-&B_{\s}^i\pa^\a\pa_\m B^{\m i}+B^{\m i}\pa_\s\pa^\a  B_{\m}
^i-\pa_\s B^{\m i}\pa_\m  B^{\a i})+(a-3)\pa_\m B^{\a i}\pa^\m  B_{\s }^i
\nonumber\\
&+&
(3a-9)\pa_\s B^{\m i}\pa^\a  B_{ \m }^i\Big)\Big].
\label{N1T1}
\eea
Their sum, that is the
complete divergent part due to the $\,\rm U(1)_Y-SU(2)_L\,$ gauge boson 
interaction is\be \G^{(1)}=\frac{4}{3(4\pi)^2\epsilon}g^\prime g^2
\kappa_2(3-a)\theta^{\m\n}\int d^4 x\,
\big(\frac 14 f_{\m\n}B_{\r\s }^i B^{\r\s i} - f_{\m\r}B_{\n\s }^i B^{\r\s i}
 \big).
 \label{G1}
\ee

Adding to this expression the divergencies which come from the commutative part 
ofthe action, and also those induced by the $\,\rm U(1)_Y-SU(3)_C\,$ mixing, we 
obtain the full result for the divergent one-loop effective action
 linear  in $\theta$:
\bea
\G_{div}&=& \frac {11}{3 (4\pi)^2\epsilon}\int d^4 x B_{\m\n}^iB^{\m\n i} +
\frac {11}{2 (4\pi)^2\epsilon}\int d^4 x G_{\m\n}^aG^{\m\n a}
\nonumber \\
&+&
\frac{4}{3(4\pi)^2\epsilon}g^\prime g^2  \kappa_2(3-a)\theta^{\m\n}\int d^4 x
\big(\frac 14 f_{\m\n}B_{\r\s }^i B^{\r\s i} - f_{\m\r}B_{\n\s }^i B^{\r\s i}
 \big)
 \nonumber \\
&+& \frac{6}{3(4\pi)^2\epsilon}g^\prime g^2_S \kappa_3(3-a)\theta^{\m\n}\int d^4 
x\big(\frac 14 f_{\m\n}G_{\r\s }^a G^{\r\s a} - f_{\m\r}G_{\n\s }^a G^{\r\s a}
 \big).
 \label{div}
 \eea
The divergent contribution due to $\rm U(1)_Y$ solely vanishes, both
the commutative and the noncommutative one.

\subsection{Counterterms}

It is clear from (\ref{div})  that the divergencies in the
noncommutative sector vanish for the choice $a=3$. Therefore one
obtains that the noncommutative gauge sector interaction is not only
renormalizable but finite. The renormalization is performed by
adding  counterterms to the Lagrangian. We obtain
 \bea \cl +\cl_{ct}&=&
-\frac{1}{4}{f_0}_{ \m\n }{f_0}^{\m\n }-\frac{1}{4}{B_0}_{ \m\n
}^i{B_0}^{\m\n i} -\frac{1}{4}{G_0}_{ \m\n }^a{G_0}^{\m\n a}
\nonumber\\
&+&
g^{\prime 3}\kappa_1\theta^{\m\n} \left( \frac 34 
{f_0}_{\m\n}{f_0}_{\r\s}{f_0}^{\r\s}-{f_0}_{\m\r}{f_0}_{\n\s}{f_0}^{\r\s}\right)
\nonumber\\
&+&
g_0^\prime g_0^2  \kappa_2 \theta^{\m\n}
\left(\frac 34 {f_0}_{\m\n}{B_0}_{\r\s }^i B_0^{\r\s i}
-{f_0}_{\m\r}{B_0}_{\n\s }^i B_0^{\r\s i}+c.p.\right)
 \nonumber \\
&+&
g_0^\prime (g_{S})_0^2  \kappa_3 \theta^{\m\n}
\left(\frac 34 {f_0}_{\m\n}{G_0}_{\r\s }^a G_0^{\r\s a}-{f_0}_{\m\r}{G_0}_{\n\s
}^a G_0^{\r\s a}+c.p.\right),
\label{lct}
\eea
where the bare quantities are given as follows:
\bea
{\ca_0}^{\m}&=&\ca^{\m }\, ,\qquad g_0 ^\prime = g ^\prime\, ,
\label{A0}\\
{B_0}^{\m i}&=&B^{\m i}\sqrt{1+\frac{44g^2}{3(4\pi)^2\epsilon}}\, ,\quad g_0 
=\frac{g\,\m^{\e/2}}{\sqrt{1+\frac{44g^2}{3(4\pi)^2\epsilon}}}\, ,
\label{B0}\\
{G_0}^{\m a}&=&G^{\m 
a}\sqrt{1+\frac{22g_S^2}{(4\pi)^2\epsilon}}\,,\quad{(g_{S})}_0=\frac{g_S\,\m^{\e
/2}}{\sqrt{1+\frac{22g_S^2}{(4\pi)^2\epsilon}}}\,  .\label{G0}
 \eea

In order to keep the constants $\kappa_1$, $\kappa_2$ and $\kappa_3$ in 
(\ref{lct}) unchanged under the renormalization procedure, i.e.
\bea
 \kappa_1=\,{(\kappa_{1})}_0\,, \:\:\;\kappa_2=\,{(\kappa_{2})}_0\,,
 \:\:\;\kappa_3=\,{(\kappa_{3})}_0\,,
 \label{kappaR}
 \eea
 we obtain the following renormalization
of the constants $C(\car)$
 \bea C_1&=&{(C_{1})}_0+\frac{33}{18(4\pi)^2\epsilon}\,,
 \quad
 %\nonumber\\
 C_2= {(C_{2})}_0-\frac{11}{18(4\pi)^2\epsilon}\,,
 \quad
 %\nonumber\\
 C_3={(C_{3})}_0-\frac{11}{18(4\pi)^2\epsilon}\,,
 \nonumber\\[10pt]
 C_4&=&{(C_{4})}_0-\frac{143}{18(4\pi)^2\epsilon}\,,
 \quad
 %\nonumber\\
 C_5={(C_{5})}_0-\frac{121}{18(4\pi)^2\epsilon}\,,
 \quad
 %\nonumber\\
 C_6={(C_{6})}_0+\frac{110}{18(4\pi)^2\epsilon}\,.
 \nonumber\\
 \label{Ci}
 \eea

Finally, an important
point is that the noncommutativity parameter $\theta$ need not  be renormalized.

\section{Discussion and conclusion}

We have constructed  a version of the standard model on
the noncommutative Minkowski
space which is one-loop renormalizable and finite in the gauge sector
and in first order in the $\theta$ parameter.
The renormalizability in the model was
obtained by choosing  six particle representations of the matter fields for the
first generation of the SM as in Table 1, and by fixing the arbitrariness in
the $\theta$-linear expansion terms in the Seiberg-Witten map.

The one-loop renormalizability
of the NCSM gauge sector is certainly a very
encouraging result from both  theoretical
and  experimental  perspectives.
So far, this property has not concerned fermions:
the results on the renormalizability of noncommutative  theories  including the
Dirac fermions are negative, \cite{Wulkenhaar:2001sq,Maja}.
However, the present result  could be an indication that the
inclusion of fermions in a renormalizable theory
might be possible by a more careful choice of representation as well.

Our result also has an important consequence on the phenomenological analysis of
the $1 \to 2$ \cite{Goran,Josip,Peter}
and $2 \to 2$ \cite{Hewett:2000zp,Ohl:2004tn,Abbiendi:2003wv}
processes in elementary particle physics. Namely, in the gauge sector of
the noncommutative standard model
the above transitions contain triple gauge boson interactions
induced by noncommutativity and, according to  (\ref{div}), they can be safely 
used further on. Since the triple gauge boson couplings have already been used
in a number of phenomenological
predictions to determine of the scale of noncommutativity
\cite{Goran,Blazenka,Josip,Peter},
the regular behaviour of these TGB interactions with respect to the one-loop 
renormalizability puts all of our predictions
from the NCSM gauge sector to a much firmer ground.

Experimentally, there are chances to detect,
in the forthcoming experiments at LHC, the decays forbidden in the SM but 
kinematically allowed, like $ Z \to \gamma\gamma$, and/or to find deviations of
${\bar f} f \to \gamma\gamma$, etc. scatterings with respect to the standard 
model predictions. Finally, the discovery of forbidden decays, and/or 
measurements of differential cross section distributions deviating from the SM 
predictions, would certainly prove a violation of the SM as we know it at 
present and could serve as a possible indication/signal for space-time 
noncommutativity.

\acknowledgments

The authors want to thank J. Wess  and
J. Louis for fruitfull discussions and the hospitality at DESY, Hamburg.
The work of V.R.  and M.B. is supported in part by the project 141036 of the 
Serbian Ministryof Science and by the DAAD grant A/06/05574. J.T. wants to 
acknowledge the support fromA. von Humboldt Stiftung as well as from the 
Croatian Ministry of Science.

\end{document}